# An authentication model based on cryptography

Ahmed AL-Arjan, Rula AL-Amrat, Rathaa AL-Ajmi

Master Students
The Hashemite University
Zarqa, Jordan
ahmed_al_argan@yahoo.com
rula.amarat@ gmail.com
Aou62017@outlook.com

Ibrahim Moh'd Salem Obeidat

Professor
The Hashemite University
Zarqa, Jordan
imsobeidat@hu.edu.jo

**Abstract**— In this paper we proposed an authentication technique based on the user cards, to improve the authentication process in systems that allows remote access for the users, and raise the security rate during exchange their messages. in this technique the server performs two functions, first function, register the users, and give him user ID, PIN code, and user private card contains secrecy information, which is used to encrypt user messages by using two kinds of encryption symmetric using RC4-Pr and asymmetric using RSA encryption., second function, distribute the user's public card if the user demand that ,in which the user sends the own authentication code with their own user ID and recipient user ID to the authentication check, and then the server sends the user public card to the recipient user, thus the sender user can send the messages to recipient user without back to the server again. We attained confidentiality using RC4-Pr and RSA encryption and message authentication, user signature, and mutual secret key by using RSA encryption.in this paper we also implement the proposal in [1] RC4-pr algorithm which is modified to improve the key weakness of basic RC4.

## 1.Introduction:

The technology in the world rapidly developed, and it becomes depends on open network systems, thus the information transmitted becomes faster between the network users, thereby we must secure the transmitted data via secured network channels by using many different techniques such as authentication factors and cryptographic, that for authenticate the users and secured the transmitted data via networks channels.

### A. Cryptography overview:

In open network systems, the information sent and received has become more vulnerable to attack by unauthorized parties, through various levels of communication, and data encryption has become the most effective of counteracting the attack [1,2,3,4].
The data encryption methods used in this field are divided into:

1. Symmetric data encryption relies on a single encryption key for encryption and decryption.
2. Asymmetric data encryption that uses public and private keys.
Public key encryption is about 1000 times slower than private key encryption, so reliance on symmetric encryption remains to encrypt all types of data.
A symmetric encryption scheme has five ingredients:
1.Plaintext: is ordinary message or data get it into an encryption algorithm to encrypt it.
2.Encryption Algorithm: is a mechanism that performs to transform the Plaintext to none understood format, depending on the secret key for performing that.
3.Secret Key: is data at a specific length that gets into an encryption algorithm with the plaintext to performs the encryption process, in which the secret key is independent of plaintext, in which the secret key is not related to plaintext.
4.Ciphertext: is an encrypted message in none understood format, which is an output of encryption algorithm, after execute some transformations and substitutions on the plaintext depends on the secret key.
 5.Decryption Algorithm, is the inverse process of encryption algorithm, that transforms the Ciphertext to the plaintext depends on a secret key as the understood format can the receiver read it clearly [5].

*1. RC4 CIPHER ALGORITHM (STREAMS CIPHER):*
RC4 is an abbreviation of the RIVEST Cipher 4 algorithm, and it is considered from symmetric cipher, which is derived from RSA data Inc, and RC4 algorithm is working by combines between the plain text and the keystream twice in the encryption and decryption process identically in each encrypts and decrypts the message.
Generally, symmetric cryptography is using one secret key for encryption and decryption, in which the sender exchanges the secret key to the recipient securely, and the secret key must be kept secrecy.
The symmetric cryptosystem has two kinds of encryption block and stream cipher, RC4 is a stream cipher, that performs to encrypt and decrypt bit by bit or byte by byte streamy, but in the block cipher will be split the plaintext into in fixed-size block, where encrypt and decrypt block by block [6,7,8,9,10,11].

The algorithm is alienated into two parts (KSA -

Key Scheduling Algorithm and PRGA - Pseudo-Random Generation Algorithm).

The initial stage is KSA, where the key K is used to permute the N-bit state table S. The KSA pseudo-code could be implemented.

In the previously mentioned procedure, bits positions in the state table S are swapped, then used as input values for the PRGA, and use the pseudorandom permutation to produce a stream of pseudorandom values. The pseudo-code of PRGA could be implemented as well [12,13,14,15].

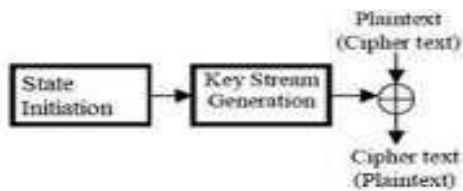

**Figure 1** Functional steps in RC4

On the sender side, the output stream Z is XOR-ed with the input stream to gain the ciphertext as in the following equation:

$$C = M \text{ XOR } Z$$

*M: input message with N bits.*

On the receiver side, the output stream Z is XOR-ed back with the ciphertext to obtain the plain text.

$$M = C \text{ XOR } Z$$

RC4 has Several softness that can be recognized. KSA is a very well-known softness of RC4, which makes the state table permuted via the fragile keys unsafe. To reinforce the KSA, the proposed work introduces a novel methodology that produces the state table of KSA in diverse mechanisms [16,17,18,19,20,21].

```
FOR i = 0 TO 255
   S[ i ] = i
   j=(j + S[ i ] + K[ i ]) mod N
   Swap(S[ i ] ,S[ j ])
NEXT
```

**Figure 2** Key Scheduling Algorithm (KSA).

```
i,j = 0
while (true)
    i=(i+1) mod N
    j=(j+S[i]) mod N
    Swap(S[ i ] , S[ j ])
    Z=S[S[i]+S[j]] mod N]
```

**Figure 3.** Pseudo-Random Generation Algorithm (PRGA).

### PREVIOUS ATTACKS ON RC4:

Due to the huge effective key of RC4, attacking the PRGA seems to be infeasible (the best-known attack on this part requires time that exceeds 2700). The only practical results related to the PRGA deal with the construction of distinguishers. Fluhrer and McGrew described in [FM00] how to distinguish RC4 outputs from random strings with 230 data. A better distinguisher which requires 28 data was described by Mantin and Shamir in [MS01]. However, this distinguisher could only be used to mount a partial attack on RC4 in broadcast applications. The fact that the initialization of RC4 is very simple stimulated considerable research on this mechanism of RC4. In particular, Roos discovered in [Roo95] a class of weak keys that reduces their effective size by five bits, and Grosul and Wallach showed in [GW00] that for large keys whose size is close to N-words, RC4 is vulnerable to a related key attack [22,23,24,25,26].

### 2. RSA CIPHER ALGORITHM (Public key encryption):

The RSA algorithm is one of the most widely accepted and implemented algorithms, it was developed in 1977 by Ron Rivest, Uday Shamir, and Lyn Adelman at MIT and was first published in 1978. The RSA algorithm consists of two keys, public and private keys, and the private key is used for decryption. To perform encryption through the RSA algorithm, we must first generate the public and private key for it, through the following steps [32]:

1- Select two prime numbers p, q
2- Compute **n=p*q**
3- Compute **Ø(n)=(p-1)(q-1)** ; where Ø(n) is Euler's totient function.
4- Choose e , so that $1 < e < Ø(n)$ , and gcd(e, Ø(n))=1
5- Determine the **Public Key (e,n)**
6- To determine the **Private Key (d,n)** must calculate d , where **d= e$^{-1}$ mod Ø(n)**

**Figure 4** Generate keys in RSA algorithm

After completing the generation of keys, encryption and decryption are performed through the following equations:

To encrypt we use the equation:

$$C = m^e \bmod n$$

And for decrypt, we use the equation:

$$M = c^d \bmod n$$

Asymmetric encryption algorithms overcome the keys distribution problem, and the strength of the RSA algorithm depends on the key length and parameter value n, so it is considered safe in the presence of large random values (p, q). But the problem with this algorithm is that the processing speed required for complex arithmetic operations, large key generation, and encryption of large texts takes a long time to do

[33].

The nature of the RSA algorithm makes it unable to encrypt large amounts of data and is sufficient for limited amounts [34], through experiments conducted by researchers to analyze the performance of the algorithms, they concluded that the RSA algorithm requires more processing time and uses more memory [35].

Therefore, they are used to encrypt the symmetric key for secure transmission.

According to many researches, the RSA algorithm is one of the most secure algorithms that have not been broken, because only the authorized person can decrypt through the private key, which remains secret to its owner [33][36][37].

To take advantage of the previous characteristics of the RSA algorithm and reduce its difficulties, many researchers resort to integrating symmetric encryption algorithms with the RSA algorithm to achieve better results in security and time spent on encryption, For example, the RSA algorithm was used to encrypt different sizes of data, and it was noted that it takes a lot of time to encrypt the larger the data size, the RC5 algorithm was used to encrypt the data, and the RSA algorithm was used to encrypt only the confidential information of RC5[38].

Also, researchers proposed an approach (Bigcrypt) based on a very good privacy policy method for encrypting big data using hybrid encryption between RSA and AES, and then testing the model on three different platforms, to obtain high security and overcome the difficulty of encrypting large data for RSA [34].

**B. Networks Attacks:**

The networks attacks are many, and the masquerade attacks are the most publication of them, which is kind of attack may be either passive such as eavesdropper attack or active attacks such as replay, interception, interruption, or modification attacks, thus the data is suffered by these attacks, thereby the triangle of security is suffered, which is represented by the confidentiality, integrity, and availability of information (CIA), the following shows the significant attacks on the networks:

*1. Masquerade Attack:*

Masquerade attack as known as Impersonation attack, and depend on to execute that by identifiers compromised to logging into IoT network, thus the network became vulnerable, In which the attacker use fake identification to authorize himself as an ordinary user, by using stolen username and password, thus the adversary becomes an authorized party in the network, and this crime depends on authentication levels in the network, the catastrophic if the criminal has full authorization, this leads to cybercrime opportunities.

In this case, the organizations must be to review authentication and authorization security in their own networks, to defend themselves from any impersonation attack in the future [27].

| Masquerade Attacks | Description |
|---|---|
| Impersonation | The attacker successfully to suppose the identity of the legitimate user. |
| Anonymity | The attacker hides their identity and performs attacks anonymously. |
| User Tracing | The attacker steals user information by track user footsteps. |
| Cloning | The attacker creates an instance of the legitimate user. |
| Identity Theft | The attacker steals the identification of the genuine user and performs malicious tasks. |
| Insider | The attacker is an authorized party that performs malicious tasks inside the network. |
| Stolen Verifier | The intruder in this attack steals verification data of current or past the authentication session from the server side and tries to get the server by using compromised data. |

| Activity Tracking | The attacker monitors the activity of legal users. |
|---|---|
| By-Passing | the attacker captures a packet from the user and responds to the user as a genuine receiving node. |

**Table 1** Description of different types of masquerade attacks:

## 2. Man-in-the-middle Attack:

Attackers secretly relay and possibly alter the communication between two parties who believe they are directly communicating with each other, and this form of eavesdropping that the eavesdropper control everything sent aside to other, which can steal valuable data such as credit card number or else, the follows figure shown how eavesdroppers perform that: [28,29].

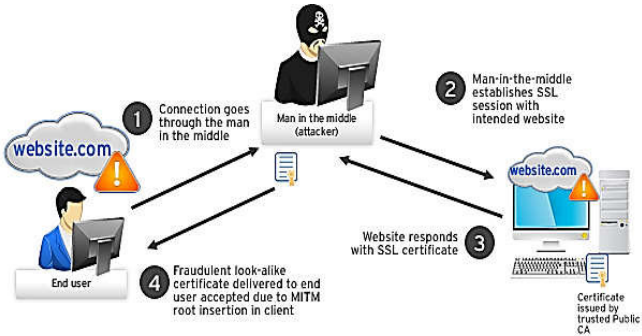

**Figure 5** Man-in-the-middle Attack

From kinds of MITM attack is ARP poison attack, in which the attacker sends forged ARP for changing mapping IP to the given MAC address of the network resource, just mapped once the cache has been modified, actually the MAC address not changed, and the attacker can act as a man-in-the-middle between two hosts in the range of the domain, and the control at the broadcasting between twice hosts [29].

## 3. Forging Attack:

In this attack, the opponent is stealing authentication info, and he got an authentication as a genuine user, and he is not, in which the forgery attacks have many kinds such that, gateway forgery, sensor forgery, Sybil attack, replay attack, and more kinds [27].

## 4. Physical Attack:

In this attack, the adversary access to IoT devices physically as an authenticated user, which most of the static devices don't have full security in the networks, thus the attack is done in different ways, such as, mobile device Loss attack, stolen card attack, stolen device attack, USB attack, and same-type-device attack [27].

## 5. Routing Attack:

In this attack, the adversary has inserted a none illegitimate node into the network, and during it convert all packets in the network to improper destination, which is done by either changing the final address of packets or sending the data packets to the wrong hop in the routing path [27].

## 6. Guessing Attack:

In cloud computing of IoT, there is an authentication mechanism done by the authentication server, which has authentication information such as, devices ID, users passwords, device secret key, thus adversaries try to getting authentication info from server to access system if the access is done, the adversary can extract this info from the server directly, otherwise if the adversary can't access to the server, in this case, he is trying to guess of passwords to authenticate themselves as a legitimate user[27].

## 7. DOS Attack:

DOS (Denial of Service) attack is a significant problem in the world of the internet, which is the target of this attack disrupts the network resources, and a network incapable to introduce the normal services, by limiting access to either machines or services rather than destroy the service itself [30].

## 8. Sybil attack:

This attack is known as also Sybil or impersonation attacks, and it's a kind of forgery attacks, In which the attacker masquerades as an authorized party and ordinary user, but in fact he's not, and use forged devices to decrease networks performance by creating passage tight for network traffic, thus the availability of network services and devices suffer.

Sybil attack can be happened by one malicious node, which broadcasting data with more than one identity, hence it leads to improper network, here very important to stay the IoT work correctly, by using strong security defense to prevent it [28,31].

In this paper, we will be shown the modified proposed of the basic RC4 algorithm as shown in [1], and we proposed an authentication scheme based on users cards such as private and public cards, which are distributed by the authentication server, After the user registration in the server, after that, the server send the user encrypted private card, in which is each card contains user information, which is saved the secrecy data for the user in own private cards such as secret key, PIN code, Authentication code, public key, and private key, and saved the public users information in the public cards such as user ID, and public key and the users can mutual the own public cards by the server, in which are used these cards for sending messages between the users themselves without back to the server again, and we used in the proposed scheme the symmetric cipher for users messages, and the asymmetric cipher for the mutual secret key, user signature, and message authentication.

## 2. The research problem

In this paper, we propose an authentication scheme based on user's cards, and we connect our proposed with the proposed that exist in [1], we looking in this paper to relay two problems:
1. First problem: The key weakness of basic RC4[1].
2. Second problem: The complicated technique of user's authentication and secure user's messages.

The first problem was solved in research [1], in which the researchers proposed to decrease the bytes in each round to 16 bytes, in addition, they proposed to add a permutation function, which is used to change the key after each encryption round by this function.

The second problem, we proposed an authentication scheme based on user cards, in which the server has two functions:
1. The first function, register the user and give her/his user ID, PIN code, and user private card contains secrecy information, which is used to encrypt user messages by using two kinds of encryption symmetric using RC4-Pr and asymmetric using RSA encryption.
2. The second function, distribute the user's public card if the user demand that, in which the user sends the own authentication code with their own user ID and recipient user ID to the authentication check, and then the server sends the user public card to the recipient user, thus the sender user can send the messages to recipient user without back to the server again.

We proposed attained confidentiality by using RC4-Pr and RSA encryption and message authentication, user signature, and mutual secret key by using RSA encryption, and we will show that in the methodology section.

## 3. Related works

### 1. Programmatically implementation of basic RC4:

The RC4 algorithm is one of the algorithms used to secure data and information and to facilitate the understanding and learning of this algorithm, and there are many previous works in which simulations of the working principle of the RC4 algorithm were presented.

In [39] the researchers demonstrated the behaviour of the RC4 algorithm and the programming of the previously described code using the programming language visual basic .net 2008, showing the stages that the algorithm goes through in the form of detailed images, showing the stage of entering the text to be encrypted and the key and converting them to ASCII code, and explaining whether the key Less than 256 bytes, the key is repeated to reach the desired limit. Then we move to the stage of creating the key used for encryption, first explaining how to create the s-box and then how to use it to configure the keystream, and finally, a screen showing the process of encrypting a text using XOR with Keystream to get the ciphertext.

### 2. The weakness of basic RC4:

More than 25 years Ron Rivest discover the RC4 stream cipher, all that time it's used widely and was one of the best achievements in the crypto word. For more than 15 years researchers know about the weakness in RC4 that help the attacker to decrypt the key stream. RC4 is a stream cipher, it Encrypt the plain text by mixing it with a series of random bytes. Making it impossible for anyone to decrypt it without having the same key used to encrypt it. but the bytes use to encrypt the plain text are not really as random as they should be at least at the beginning of the process this is the known weakness in RC4.According to that weakness the RC4 suffer from many attacks during all the previous years we will mention some of them in this paper [40].

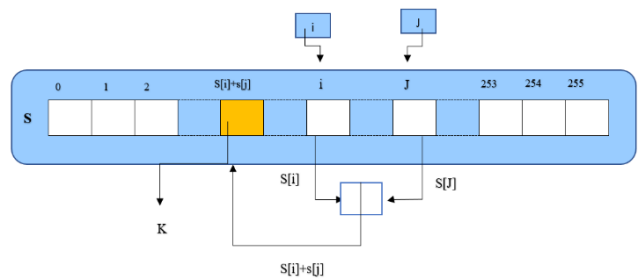

**Figure 6** RC4 stream cipher

### I. How RC4 runs:

Rc4 runs in two phase the first part is key scheduling algorithm KSA which is takes an arrays s or s-boxes to derive a permutation of $\{0,1,2\ldots\ldots, N-1\}$ using a variable size key k.

The second part is the output generation part PRGA which prod use pseudo-random bytes using permutation derived from KSA. each loop or round produce one output value .RC4 is used with word length n= 8 bits and n= 256 [41].

### II. Attacks caused by RC4 weakness:

Due to the known weakness of the RC4 algorithm, some researchers have been able to present how the algorithm can be penetrate by the attackers by exploiting the weakness. We mention some attacks below:

Knudsen et al. they attack one version of RC4 with n<8 by their backtracking algorithm in which the adversary guesses the internal state and checks if an anomaly occurs in later stage. In the case of contradiction, the algorithm backtracks through the internal state and re-guesses. This remains the only effect algorithm which attempts to discover the secret internal state of this cipher.

A serious weakness in RC4 was observed by Mantin and Shamir who noted that the probability of a zero-output byte at the second round is twice as large as expected. in broadcast applications a practical cyphertext only attack can exploit this weakness.

A probabilistic correlation between the secret information (s, j) and the public information (I, output) is discovered by Jenkins. On the other hand, Golic detected a positive correlation between the second binary derivative of the least significant bit output sequence and 1. This correlation can be distinguished from the random stream of bits by only $2.7^{44}$ outputs bytes.

Grosul and Wallach showed that a related key attack works better on a very long keys, another weakness is also discovered by Andrew which is a classis of weak keys on RC4.

Also, if some portion of the secret key is known the RC4 can be broken completely. This is what Fluhrer et al is demonstrate. This is very important discover because in the wired Equivalence privacy protocol (WEP) a fixed secret key is concatenated with IV modifiers for encrypting the messages, this is shown the attach ca be done.

Another weakness was observed recently by Paul and Preneel they designed an algorithm to deduce certain special RC4 states known as (non-fortuitous predictive state). And they proved that only a known elements of the s-box along with two index-pointers cannot more than an output byte in the next random.

Daniel J. Bernstine a professor in university of IIIinoins at Chicago presented his research on secret key cryptosystem discover a new attack on the RC4 that enable the attacker to compromise a victim's session with a site protected by TLS. that's mean attack against TLS/RC4 is possible.

### III. A new weakness in RC4:

A new weakness in the RC4 is discovered by the researchers. They observe that the distribution of the first two output bytes not uniform. This makes the RC4 trivial to distinguish between short outputs of rC4 and random string by analyzing their first or second outputs values of RC4 or Diagraph. And they note that the probability that the first two output bytes are equal. Fluhrer and McGrew showed that the first two outputs take the value (0,0). Experiments observed that this result is incorrect. [41.42]

### 3. The key vulnerability of basic RC4:

A. Mughaid, A. Al-Arjan, et al in [1], are proposing a permutation function that is used to changing the key of every stream 16 bytes encrypted even at the end of plaintext, thus the subkeys count are equal to the count of bytes divided by 16, and adding one to the integer result if remainder, not equal zero, for instance, if we need to encrypt of 321 bytes then, we need ((321/16)+1) is equal to 21 subkeys used to encrypt 321 bytes, we will illustrate this proposed in this section.

### i. RC4-Pr modified algorithm:

This paragraph explains what has been modified in the basic algorithm, to avoid any attack that depends on determining the statistical relationships between the encrypted text and the encryption key used to discover either the plaintext or secret key. Furthermore, this modification was made by doubling the cipher key scheduling in each round KSA - 2x, the first scheduling is done to generate the new subkey for a round Encryption, and the second scheduling is used to generate the stream keys to encrypt the data of this round. This process is repeated for each round of encryption to reach the last round. Also, the key length is specified as 128 bits, and the length of the data in the round is also 128 bits [1].

#### A. Generate subkeys by Pr function:

The Pr function is an importance in the modified algorithm to generate all the subkeys used in the key scheduling before the cipher stream begins in each cipher round. Where the encryption key used in this round is sent to the Pr function, that it is transposing bytes to produce a different new subkey and send it to the next round, to complete the encryption process. knowing that the working mechanism of the Pr function is similar to the work of the Key Scheduling Algorithm (KSA), but here the function performs a permutation to arrange the byte in the receiving key to generate a new subkey. Noting that the number of subkeys that canbe generated from the primary encryption key is N factorial. Where N is the length of the key suggested here, 128 bits, the following figure shows how the Pr function works [1]:

> $S=(0,1,\ldots,n-1)$ ,*where n is K length.*
>
> $K=(b_0,b_1,\ldots,b_{n-1})$,*where b's are K bytes.*
>
> *//the following FOR loop is for permutation bytes orders in S by two pointers i, j.*
>
> **For i from 0 to (n-1)**
>
>    **j=(j+s(i)+K(i)) mod (n-1)**
>
>    **swap(S(i) , S(j) )**
>
> *//the following FOR loop is for update permutation bytes orders in K by S.*
>
> **For i from 0 to (n-1)**
>
>     **swap(K(i) , K(S(i)) )**
>
> *//the following code return new round key for current block.*
>
> **Return K**

**Figure 6** Pr function implementation code

#### B. RC4-Pr implementation scheme

As previously explained, the difference between the basic algorithm and the modified algorithm is the addition of a function to generate the subkeys, which are used in all rounds of encryption, the following figure shows how the modified algorithm works:

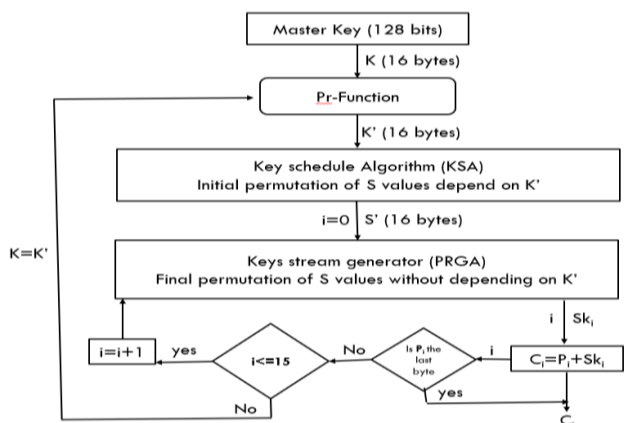

**Figure 6** RC4-Pr diagram

## 4. Methodology of propose model:
### i. *Programmatic implementation of RC4-Pr-Modified algorithm:*

we implemented programmatically the RC4-Pr, which is proposed in [1], by using python language version 3.8, which took 80 programmatically code lines, and we tested encrypt on 7 files and computed some results such as files size, encryption duration time, the number of subkeys used, and the number bytes encryption rate per millisecond, and finally, we computed total results of 7 encrypted files, the following results from python screen as it appeared after execute the program:

```
===================File # [1] ============
The file size:  1779700 Bytes
Encryption duration time:  1936.4478588104248 MS
The number of subkeys:  111232
Byte's encryption rate:  919.0539223159274 Bytes/MS
===================File # [2] ============
The file size:  15712007 Bytes
Encryption duration time:  19655.32088279724 MS
The number of subkeys:  982001
Byte's encryption rate:  799.3767740394147 Bytes/MS
===================File # [3] ============
The file size:  83987 Bytes
Encryption duration time:  19801.0196685791 MS
The number of subkeys:  5250
Byte's encryption rate:  4.2415492437125994 Bytes/MS
===================File # [4] ============
The file size:  366889 Bytes
Encryption duration time:  20207.38196372986 MS
The number of subkeys:  22931
Byte's encryption rate:  18.156186717236675 Bytes/MS
===================File # [5] ============
The file size:  25284 Bytes
Encryption duration time:  20236.438751220703 MS
The number of subkeys:  1581
Byte's encryption rate:  1.249429324538381 Bytes/MS
===================File # [6] ============
The file size:  1337344 Bytes
Encryption duration time:  21921.06032371521 MS
The number of subkeys:  83584
Byte's encryption rate:  61.00726790816774 Bytes/MS
===================File # [7] ============
The file size:  323742 Bytes
Encryption duration time:  22307.792901992798 MS
The number of subkeys:  20234
Byte's encryption rate:  14.512506971098853 Bytes/MS
===========Total Result of encrypt 7 files======
The total files size:  19628953 Bytes
The total Encryption duration time: 126065.46235084534 MS
The total number of subkeys:  1226813
     Byte's encryption rate:  155.7044461977367 Bytes/MS
```

**Table 2** The results of encrypt 7 files by RC4-Pr

### 1. RC4-Pr time efficiency:

From the above results, we notice the program needs 126065.4 milliseconds to encrypt 19628953 bytes, which is used 1226813 subkeys at a length of 128-bits for each subkey, and the bytes encryption rate was 155.7 bytes/milliseconds, in another means the program needs 1 minute to encrypt 9 megabytes, thereby we consider the time efficiency is accepted for encrypting/decrypting authentication messages and user's messages in our proposed.

### 2. Users' authentication technique model:

In this paper, we proposed an authentication model based on cryptography, and we determined two cryptographic algorithms for use in authentication operations either between the user and server or between the user and another user, as further mention, we will use the RC4-Pr algorithm that proposed in [1], and we will use RSA encryption, and our model divided into two phases:

### A. The User registration phase:

In this section, we show the steps of the user registration by connecting with the authentication server and getting the user private card at the end of the registration phase, the following steps show user registration by the authentication server:

**Step1:** in this step, the novel user connects with the authentication server and requests the registration, in which the server request the novel user to enter a valid email address and mobile number.

**Step2:** in this step, the novel user will input each of his/her email address and mobile number and send them to the server.

**Step3:** in this step, the server receives the user email and mobile number without any encryption, and the server will generate unique user-id and PIN-code and encrypts them by its own private key and send it to the received email, and the server sends its own public-key via SMS message to the received mobile number.

**Step4:** in this step, the user will receive the encrypted authentication message via its own email, and at the same time the user received via SMS the server public-key, and the user will decrypt the message by received public-key and extract user_id, PIN-code, that his/her uses the extracted information to compute its own authentication code.

**Step5:** in this step, the user computes its own authentication code by the following formula:

$$\text{Outh-code} = h(uid \| upn \| uml \| umn) \ldots 1$$

Where:
- outh-code: is the user authentication code.
- h: is a hash function.
- uid: is the user id.
- upn: is the user PIN-Code.
- uml: is the user email address.
- umn: is the user mobile number.

**Step6:** in this step, after the user computed its own outh-code in the previous step, in which the user encrypting the user-id and outh-code by using the RC4-Pr algorithm and send the cipher message to the server, the following details show how to create the user secret key and used it for encryption operation:

1. *Create user secret key:*

$$K = Md5\_h(uid \| upn) \ldots 2$$

Where:
- K: is the user secret key.

Md5_h: is Md5 hash function.
uid: is the user id.
upn: is the user PIN-Code.

## 2. *Encrypt user authentication message:*

To encrypt the user authentication message by RC4-Pr we need a secret key at a length of 128-bits, this illustrates why we used Md5 hash function because the result of the Md5 hash function is a fixed block at the length of 128 bits, thereby we used it as a secret key to RC4-Pr algorithm [1], the following show the encryption line function:

$$M= (uid||outh\text{-}code|| T_1) \ldots\ldots 3$$
$$M=M||h(M) \ldots\ldots\ldots\ldots\ldots 4$$

Where:
M: is the user authentication message.
outh-code: is the user authentication code.
uid: is the user id.
$T_1$: is the user current timestamp.
h(M): is hash value of **M**.

$$C=E (M, K) \ldots 5$$

Where:
C: is the ciphertext of message.
E: RC4-Pr encryption function.
M: is the user authentication message.
K: is a secret key created by formula 2.

**Step7:** in this step, the server receives an encrypted authentication message from the user, and computes the secret key by using formula 2 and decrypts the received message using the computed secret key, and extracts the user id, outh-code, $T_1$, h(M).

**Step8:** in this step, the server computes the server current timestamp ($T_C$), and computes ΔT by using $T_C$ and received $T_1$, by using the following formula:

$$\Delta T= T_C - T_1 \ldots 6$$

**Step9:** in this step, if the ΔT computed value by formula 6 is less than or equal allowed delay time range, then the server move to check the received message authentication by computing the hash value of the received message after extract the received hash h(M) from the message, and the server compare the computed hash value with the received hash value if they equal then, the server creates the user's private card and encrypts it by RC4-Pr using the user secret key computed by formula 2, and send it to the email address that received from the user.

**Step10:** in this step, if either the ΔT computed value by formula 6 is greater than the allowed delay time range, or the computed hash value not equal to the received h(M) value then, the server no does any further action and sends a rejection message by email address to the user.

## 3. *The user cards:*
### 1. The Private card:

The user private card is a card that contains the user secret information that her/his uses for connecting either with the server or another user and this information is used to encrypt the sent messages from the user, the following table mention the user private card contents:

| Field | Description |
|---|---|
| User_ID | The user identifier |
| User_Mobile_Number | The user mobile number. |
| User_Email | The user email address |
| Outh_code | The user authentication code |
| PIN-code | The user PIN-Code, random 6 mixed digits and letters. |
| S-key | The user secret key |
| Server-Pub | The public key of the authentication server |
| Pr-key | The user private key |
| Pub-key | The user public key |

**Table 3** The user private card

### 2. The Public card:

The user public card is a card that contains the public information of the user, which the user sends to another user for connecting with each, and the exchange public cards are the responsibility of the server after the users received the public card, which they can connect with each other without back to the authenticator server, the following table mentions the user public card contents:

| Field | Description |
|---|---|
| User_ID | The user identifier |
| User_Name | The user's real name |
| User_Mobile_Number | The user mobile number. |
| User_Email | The user email address |
| Pub-key | The user public key |

**Table 4** The user public card

## B. The User Logging phase:

In this section, we illustrate the steps that the user needs to login into the system:

### *i. The User Logging phase steps:*

**Step1:** in this step, the user asks the server to log in to the system by sending the user his user-id.

**Step2:** In this step, the server searches its black list of users, if the user is in the list, the server rejects this request and sends a rejection message to the user via his/her email.

**Step3:** in this step, if the user that requests to login does not exist in the block list then, the server request from the user enters the user-id and PIN-code.

**Step4:** in this step, the user extracts the outh-code from his private card after decrypting it and creates an authentication message (M) by formulas 3 & 4, and encrypted by server public key, and sends it to the server, the following formula shows the user authentication message encryption:

$$Auth\_Msg= (M, pub\_key) \ldots 7$$

Where:
M: is the user authentication message.
pub_key: The server public key

**Step5:** in this step, the server received the user encrypted Auth_Msg and decrypting it by using its own private key and extract the user-id, outh-code, T1, h(M), the server computes the server current timestamp (TC) and computes ΔT by using formula 6.

**Step6:** in this step, if the ΔT computed value by formula 6 is less than or equal allowed delay time range, then the server move to check the received message authentication by computing the hash value of the received message after extract the received hash h(M) from the message, and the server compare the computed hash value with the received hash value if they equal then, the server generates OTP-code and send it via user SMS, and requests from the user to enter it.

**step7:** the server checks OTP-code is equal with saved OTP-code then the user enters to its own dashboard.

**step8:** if each of ΔT, h(M), and OTP-code have not attained server conditions then, the server no does any further action and sends a rejection message by email address to the user.

*ii. Block user account:*

The user prevents logging in if only if fails three times to login into the system, which the server adds the user to black list and block user account until, the server sends an email message contains sensitive data about the device that tries to log in, such as, device name, device IP address, Date-Time, and information of Internet service provider for this device, in addition, the server generate activation code and sends it to the origin user via SMS by registered mobile number, in which the server asked from origin user to enter the received code to activate his/her account then, if the user entered correct code then, the server allows to user to log in again.

*iii. User dashboard:*

After the user login successfully appears user dashboard, which is used for user procedures such as, view his/her private card, sends his/her public card, sends a message to another user, and view the publics cards list that is saved.

**1. View user private card:**

The user can view his private card after login, that which contains user secret information used in connection operations between the user and others or with the authentication server, especially it used to encrypt/decrypt the received/sent data, the following shows user13 private card after decrypted it by RC4-Pr algorithm by using user secret key that computed by formula 2:

```
User_Id ====> user13
outh_code ====> 3076315706752804757
pin_code ====> ZTMw9G
S_Key ====>
48_102_192_245_224_165_19_50_81_78_15_225_
56_140_76_105
server_pub_key ====> 21883-5
private_key ====> 75137-31963
public_key ====> 75137-7
```

**Figure 7 User private card**

**2. Send user public card to another user:**

The user can send his public card to another user, in order to connect with him, which is contains the public information of the user, such as, user id, name, mobile number, email address, and the user public key, which the user sends its own user id, recipient user id, and its own the outh_code to the server, the following steps show how to the user can send public card:

**step1:** the user asked the server to sends his public card to the recipient user.

**step2:** the server asked the user to send the sender id, the sender outh_code, and the recipient id.

**step3:** the user creates an encrypted message by the server public key, which the encrypted message is contained the user id, outh_code, and the recipient id, and send it to the server.

**step4:** the server decrypts the received message by its own private key and extracted each sender id, sender outh_code, and recipient id, in which the server checks of extracted data if it is correct then, the server sends the user public card to the recipient via his email.

**step5:** if extracted data in step4 is not correct then, the server sends a rejection message to the user via email.

**3. Send messages:**

The user can send a message to another user without connecting with the server to do that, but the sender user must have the public card of the recipient user, because the card contains important information used in sending, such as user id, email address, and public key of the recipient, the following steps show how to the user can send the message to another user:

*1. Sending message:*

**step1:** the user decrypts his private card and extracts a secret key (K), and his private key then, the user extracts the email address and the public key from the recipient's public card.

**step2:** the user computes the hash value h(msg) of the message before encrypting it, and computes the current timestamp ($T_1$).

**step3:** the user creates an authentication detail (**auth-dtls**), which is added to the end of the encrypted message, the following formula shows that:

$$\text{u-signature} = E\ (user\_id \| h(msg) \| K \| T_1, PR_{Sender}) \ldots 8$$
$$\text{auth-dtl} = E\ (u\text{-signature}, Pub_{Recipient}) \ldots 9$$

**step4:** the user computes a secret key for the recipient (K') by using the recipient id and sender secret key that is used to encrypt/decrypt the message, the following formula shows how the user computes K':

$$K' = h\ (Recipient\text{-}ID \| K) \ldots 10$$

**Step5:** the user encrypts the message (msg) by using RC4-Pr by using his secret key (**K'**), by the following formula:

$$Cmsg = E\ (msg, K') \ldots 11$$

**Step6:** the user adds auth-dtl that computed by formula 8 & 9 to the end of Cmsg that encrypted by formula 11 as one message and sends it to the recipient email.

*2. Receiving message:*

**step1:** the recipient user as an initial procedure separates each of the received encrypted the message into Cmsg that encrypted in formula 10 and auth-block that created by formula 8 & 9.

**step2:** the recipient user decrypts auth-dtl twice, the first one ($D_1$) by using its own private key, and the second one ($D_2$) is decrypted ($D_1$) by using the sender public key, the following formulas show how to do:

$$D_1 = D\ (auth\text{-}dtl, PR_{Recipient}) \ldots 12$$

$$D_2 = D(D_1, Pub_{Sender}) \ldots\ldots\ldots 13$$

**step3:** if $D_2$ is readable format then, the recipient extracts each of sender id, h(msg), K', and $T_1$.

**step4:** the recipient is decrypting extracted Cmsg by using extracted K' and computes the hash value of the decrypted message h'(msg), and compares extracted h(msg) with h'(msg), if they are equal then the message authenticated and it was not altered, otherwise, the recipient sends a rejection message to the sender via email.

### 4. View public cards list:
The user can view all public cards that are saved, which each card contains the public information of a certain user as mentioned previously in Table 4.

### C. Request secret information phase:
In this section, we show how the user can request his own secret information, such as PIN-code, private card, and server public key, furthermore, each user has a different server public key, thus just a certain user can decrypt server message that encrypted by server private key that custom for this user.

#### 1. Request server public key:
The server in the registration phase generates the private and public keys for a novel user, and at the same time generates its own private and public keys for a special connection with this user, which the server has different private and public keys for each user individually, the following steps show how a user can request its own server public key:

**step1:** the user asked the server to get its own server public key.
**step2:** the server asked the user to enter his user-id.
**step3:** the user enters his own user-id and sends it to the server.
**step4:** the server checks of user-id and send his own server public key via SMS by registered mobile number in the server.

#### 2. Request PIN-code:
##### 2.1. Request resend PIN-code:
The user can be asked the server to resend his PIN code, because the server saved the encrypted authentication message that was sent in the user registration phase via the user email, thus the server resends the message again via the same user email, by the following steps:

**step1:** the user asks the server to resend his PIN code.
**step2:** the server asks the user to enter user-id.
**setp3:** the user sends his user id to the server, and the server receives it and checks of user id, and resends the encrypted server authentication message via the user email.
**step4**: the user receives the server message, and decrypts it by the server public key, and gets his PIN code.

##### 2.2. Request change PIN-code:
The user can change his PIN code by asks the server to do that, the following steps show how the user can ask that:

**step1:** the user asks the server to change his PIN code.
**step2:** the server asks the user to sends its own user-id and outh_code.
**step3:** the user sends its own user-id and outh_code to the server.
**step4:** the server checks of user-id, and generates a new PIN code of the user, and the server computes a new outh_code by formula 1 and computes a new server authentication message of the user, and the last procedure the server updates the user private card and sends the server authentication message and updated private card to the user by his email.
**step5:** the user receives the server message by its own email, and decrypts the message by the server public key, and gets a new PIN code, and gets an updated private card.

#### 3. Request private card:
The user can ask the server to resends his private card, which the server asks the user to enter user-id and PIN code for the check then, the server resends the encrypted user private card via email.

#### 4. Request change contact information:
The user can ask the server to change its own email address or mobile number, all the user's messages encrypted by the server public key at all the user steps, by the following steps the user can change that:

**step1:** the user asks the server to change its own contact information.
**step2:** the server asks the user to sends user-id and outh_code for checks user-id.
**step3:** the user enters his own user-id and outh_code, and sends them to the server.
**step4:** the server checks of user-id, and asks the user to enters her/his email address.
**step5**: the user enters his own email address and mobile number, and sends them to the server.
**step6:** in this step, the server will be registered the user again but by the same user-id, thus the server goes to generates its own public and private keys and for the user, and user PIN code and the server will use the same steps in the user registration phase from step3 to step6.

### 5. The authentication model results:
Our proposed model achieved the highest levels of security, confidentiality, message authentication, digital signature, secure secret key exchange, and prevention of attacks. The following explains all of the above in our research paper:

| Result | Description |
|---|---|
| Confidentiality | When the user at the registration phase sends an encrypted message containing the user_id and outh-code using the RC4-Pr algorithm and the secret key to the server, only the server can decrypt it through the shared secret key. |
| | When the server creates the user's private card, then sends it encrypted to the user by the RC4-Pr algorithm and the secret |

| | | | |
|---|---|---|---|
| | | | key known only to the user and the server, only the user can decrypt this message. |
| | | | The user-sender used RC4-Pr to encrypt his message by K' secret key and encrypt an authentication detail by RSA using his private key and the recipient's public key. |
| Message authentication | | | the user encrypts authentication message (uid ‖ outh-code ‖ T1 ‖ h (M)) by the secret key, then is sent to the server, here is the only server that can decrypt it with the shared secret key, the server and the user who sent the message Only those who have the key (here check one of the authentication conditions that the message came from the alleged party), to make sure of the second condition that the message has not changed, from calculating the hash value of a message. |
| | | | When the user logs into the system, an authentication message is generated from the user and is encrypted by the server's public key and sent to the server, as shown in Equation No. 7, where the server knows who the sender is because at the beginning of the user's registration, the server generates a key Public and private for each user, and the server generates a public and private key to communicate with each user individually, and the server is the party that can decrypt the message. |
| Digital signature | | | When messages are exchanged between two users, the sending user creates a message u- |

| | |
|---|---|
| | signature which is a message encrypted with the sender's private key and used as an authentication technique to ensure that the sender does not deny or repudiate a message, which is re-encrypted using the recipient's public key, shown in equations 8 and 9. |
| Secret Key exchange | The user-sender can exchange a secret key with the recipient by using RSA encryption, which we showed in create authentication detail. |
| Attacks prevention | If the user fails to log into the system three times, the server will block the user account in anticipation, and then send an email to the original user containing sensitive information about the device that tried to access the user account. |
| | The recipient of the message can detect if somebody is in the middle by computing delay time ΔT after receiving the message. |

**Table 5** The authentication model results

## 6. Discussion:

The proposed model achieves many security aspects, such as safe the messages transmitted from passive or active attacks by using symmetric encryption, in which we used RC4-Pr to encrypt these messages, furthermore, we used asymmetric encryption to achieves safe exchanging of secret information by using RSA encryption that to resolve many important issues such as, sender digital signature, data confidentiality, message authentication using the hashing, and secret key exchange, in addition, prevent some of the attacks especially masquerading attacks by using cryptography, hashing, and timestamps, and we used multi-factor authentication by using PIN-code, OTP-code by sending them via email address and SMS of users.

## 7. Conclusion and Future work:

The aim of developing authentication techniques is to prevent illegality users by finding modern authentication techniques that depend on multi-factor authentication, which depends on cryptography during data transmitted between the nodes of the network, thus preventing many in-the-middle attacks such as masquerading attacks. User authentication is an important part of the security of systems, which the user is very interested in, thus the researchers always looking to develop it, therefore we implemented the modified RC4-pr algorithm in [1] to cover the weakness of basic RC4 in the first part of the paper. and In the second part, we improved the authentication process by designed an authentication technique with highly efficient security, in which we introduced an authentication technique based on the user cards. in this technique, the server s performs two functions, the first function pass through three stages (registration, logging, and dashboard). The second function, distribute the user's public card if the user demands that. which are both functions mention in the methodology by steps and details. In the last comment, we would like to mention the security of users could be further improved by adding an authorization card in our model in the future.